\documentclass[11pt]{article}
\usepackage{amsmath,amstext,amsbsy,amssymb}
\usepackage{bm}
\usepackage{color}

\newcommand{\Tr}{{\rm Tr}}
\newcommand{\sign}{\text{sign}}

\newcommand{\ssI}{{\scriptscriptstyle{ I}}}

\newcommand{\ssD}{{\scriptscriptstyle{ D}}}

\newcommand{\ssH}{{\scriptscriptstyle{ H}}}
\newcommand{\ssS}{{\scriptscriptstyle{ S}}}
\newcommand{\ssF}{{\scriptscriptstyle{ F}}}
\newcommand{\ssE}{{\scriptscriptstyle{ E}}}

\newcommand{\ssh}{{\scriptscriptstyle{1/2}}}

\newcommand{\ssprime}{{\scriptscriptstyle{ \prime}}}

\long\def\symbolfootnote[#1]#2{\begingroup%
	\def\thefootnote{\fnsymbol{footnote}}\footnote[#1]{#2}\endgroup}

\begin{document}
	
	\title{ Spin Currents of Charged Dirac Particles in Rotating Coordinates}
	
	\author{\"{O}. F. Dayi, E. Yunt \\
		{\em Physics Engineering Department, Faculty of Science and
			Letters, }\\ \em {Istanbul Technical University,
			TR-34469, Maslak-Istanbul, Turkey
			\footnote{{\it E-mail addresses:} dayi@itu.edu.tr,
				 yunt@itu.edu.tr } }}
	\date{}
	
	
	\maketitle
	
\begin{abstract}

The semiclassical Boltzmann transport equation of  charged, massive fermions in a rotating frame of reference, in the presence of external electromagnetic fields  is solved in the relaxation time approach to establish the distribution function up to linear order in the electric field in rotating coordinates, centrifugal force and the derivatives.  The spin and spin current  densities  are calculated  by means of this distribution function at zero temperature up to the first order.  It is shown that the nonequilibrium part of the distribution function yields the spin Hall effect  for fermions constrained to move in a plane perpendicular to the angular velocity and magnetic field. Moreover it yields an analogue of Ohm's law  for spin currents  whose  resistivity depends on the external magnetic field and the angular velocity of the rotating frame. Spin current densities  in three-dimensional systems  are also established.

\end{abstract}

\maketitle

\newpage
\section{Introduction}

In spintronics a major field of investigation is the efficient generation of spin current which is mainly achieved through the intrinsic spin-orbit and spin-magnetic field interactions. When a fermionic system is in a rotating frame of reference, spin also couples with the rotation of the system. This spin-rotation coupling  theoretically offers an alternative mechanism to generate spin currents  \cite{mism,mism2}.  It gives rise to the possibility of  generating spin current mechanically, without the limiting requirement of strong intrinsic spin-orbit coupling in condensed matter systems. Some features of the spin-rotation coupling is discussed in \cite{chowbasu} by pointing out the similarity of this interaction with the intrinsic spin-orbit and the Zeeman interactions. These  formalisms    are mainly based on the Pauli-Dirac type Hamiltonians which are suitable to study the nonrelativistic  dynamics of charge carriers. 

The Dirac equation in a noninertial frame of reference is established in \cite{HehlNi}. The electron spin couples in a similar manner  with the angular velocity of rotation $\bm\Omega$ and  the external magnetic field $\bm B.$  Furthermore, the Maxwell's equations are modified in rotating coordinates \cite{Schiff}. For nonrelativistic rotations, $|\bm \Omega \times \bm x|\ll c,$ the Maxwell's equations are
\begin{eqnarray}
\bm \nabla \cdot \bm E^\prime & = &4\pi q n,\nonumber \\
\bm \nabla \times \bm E^\prime & = &- \frac{\partial \bm B}{\partial t},\nonumber\\
\bm \nabla \cdot \bm B & = & 0, \nonumber \\
\bm \nabla \times \bm B & = &4\pi q \bm j + \frac{\partial \bm E^\prime}{\partial t}, \label{Meq}
\end{eqnarray}
where the electric field in the rotating frame is $\bm E^\prime =  \bm E + (\bm\Omega\times \bm x)\times \bm B .$ When a rotating object which possesses  fermionic charge carriers, is subjected to the external electromagnetic fields $\bm E$ and $\bm B,$ these external fields will evolve in it according to the Maxwell  equations (\ref{Meq}) where the particle
number and current densities, $n,$ and $\bm j,$ should be consistently  furnished.

Son-Yamamoto \cite{soy} and Stephanov-Yin \cite{sy} showed that  chiral anomalies can be  incorporated  into the semiclassical kinetic theory of chiral particles. Since then the semiclassical formulation has been extensively employed in studying dynamics of fermions either massless or massive.  It furnished  intuitive understandings of phenomena like the chiral magnetic  effect \cite{kmw,fkw,kz}, the chiral separation effect \cite{mz,jkr}, the chiral vortical effect \cite{ss} and local (spin) polarization effect \cite{lw,bpr,glpww}. 	

We would like to obtain the spin  and current density of Dirac particles within the semiclassical kinetic theory in the presence of the external electromagnetic fields in a uniformly rotating coordinate frame. 
The  rotations are  nonrelativistic but in contrary to Refs.\cite{mism, mism2, chowbasu} we deal with the dynamics of 
particles (antiparticles) considered  as the wave packets composed of  positive (negative) energy solutions of the free Dirac equation.
The  semiclassical method which we employ is a differential form formalism based on  these wave packets.
This system is not covariant,  although fermionic particles obey relativistic dispersion relations.

In this approach  the Berry curvature arises naturally \cite{sniu, cyniu, om-elif}  and it is incorporated in
the underlying  symplectic two-form \cite{ds, de}. The semiclassical kinetic theory of the Weyl and Dirac particles in the presence of the external electromagnetic fields in a rigidly rotating coordinate frame is elucidated in Ref. \cite{oee}. There the matrix-valued  phase space measure and the time evolution of phase space variables are obtained in rotating coordinates.

One particle dynamics can be generalized to many particles by means of the  kinetic theory. The semiclassical phase space velocities  can be employed to acquire  the related  Boltzmann transport equation whose solution will be the nonequilibrium distribution function in the presence of collisions. It is worth noting that  we do not deal with  nonequilibrium thermodynamics, we consider the nonequilibrium state of a closed system. The relaxation time approach offers an accessible  technique to consider collisions \cite{anselm}.  Chiral kinetic theory is studied  within this method in \cite{hyy} . It is  difficult to solve the transport equation on general grounds, so that   one  should resort to approximations. The distribution function can be expanded in a series of the external electric field and solved perturbatively up to the desired order. Actually, we keep terms up to linear order in the electric field  in rotating coordinates, the centrifugal force and  the  derivatives of the chemical potential.  We deal with a roughly neutral background due to the presence of particles and antiparticles. The distribution function which will be established consists of the equilibrium part chosen  to be the Fermi-Dirac distribution function and the nonequilibrium part which corresponds to the first order term. 

Spin currents are defined in terms of the distribution function,  first time derivative of spatial coordinates weighted with the measure of the phase space and  Pauli spin matrices. The corresponding spin densities are given  by the measure of  phase space and the spin matrices. Here only the zero absolute temperature is taken into account in calculating the spin and spin current densities.

We first deal with the charge carriers constrained to move in a plane perpendicular to the magnetic field  $\bm B$ and the angular velocity $\bm\Omega.$ These may have some relevance in the context of generating  spin currents in two-dimensional condensed matter systems like metal films. We show that the equilibrium distribution function generates the spin Hall effect associated to the electric field in  the rotating frame of reference, whose  conductivity  depends on  the chemical potential $\mu$ (Fermi energy)  and the mass of the Dirac particle $m. $ In the nonrelativistic   limit  it produces the spin Hall conductivity calculated in Ref.\cite{ccn} for an inertial reference frame as well as  the one obtained in Ref.\cite{mism2} for rotating coordinates in the presence of magnetic field. In the $\mu \gg m$ limit it yields  the topological spin Chern number \cite{prodan,ezawa} as it was discussed in Ref.\cite{biz-ann}. This limit actually is equivalent to consider massless case given in Refs.\cite{soy, sy}. We mainly make use of Berry gauge fields which are defined in adiabatic approximation where level crossing is not allowed. In fact when one does not allow level crossing Lorentz invariance of the system is broken even in the absence of rotation \cite{cssyy, dehz, css, hpy}.   Ref.\cite{fuj} provides insights about the role of level crossing and the adiabatic approximation in obtaining the Berry phase.
We show that the spin Hall effect associated to the electric field in rotating coordinates results also from the nonequilibrium distribution function with  the spin Hall conductivity independent of the magnetic field and  angular velocity. We also derive an Ohm's Law analogue for the spin current, where the analogue of resistivity depends on $\bm B$ and $\bm\Omega$ as well as on the chemical potential and mass. It is shown that the spin current is conserved up to first order.

When the Dirac particles are free to move in all three space dimensions, we integrate over the three dimensional momentum space to obtain the spin current densities. We study the  spin current densities in three-dimensional conductors by keeping the direction  of spin arbitrary. We  find out that to generate spin currents in a certain direction the angular velocity of rotation or the external magnetic field should possess a nonvanishing component in that direction.  The three-dimensional spin current yields similar effects with the two-dimensional system. The spin Hall conductivity arising from the nonequilibrium distribution function in three dimensions is analogously independent of the fields $\bm B$ and $\bm\Omega$ and depends only on the chemical potential and mass.

The paper is organized as follows.  An overview of the semiclassical formalism of  the Dirac particles in rotating coordinates, in the presence of electromagnetic fields is presented in Section \ref{STDP}. We present the derivation of the nonequilibrium distribution function in the relaxation time approach  in Section \ref{SCB}. The definitions of the spin  and spin current densities  for the Dirac particles are given in Section \ref{SCD}. In Section \ref{2D}, we discuss the spin and spin current densities in two-dimensional conductors. We obtain the spin Hall conductivity and an Ohm's Law analogue for spin associated to the  electromagnetic fields in rotating coordinates at $T=0.$  In Section \ref{3D}, we calculate the spin current densities in  three-dimensional conductors. In Section \ref{CONC}, we discuss our results.  

\section{Semiclassical Velocities
}
\label{STDP} 
 
We work within the  semiclassical approach   based on the wave packets composed of the positive and negative energy solutions of the free Dirac equation: 
\label{STDP} 
\begin{equation}
H_\ssD^{\scriptscriptstyle{(4)}} (\bm{p})=\beta m + \bm{\alpha} \cdot \bm{p}.
\label{hamiltonian}
\end{equation}
We set the speed of light $c=1,$ and choose the following representation of
 $\beta,\ \alpha_i;\ i=1,2,3,$  matrices,
$$
\bm \alpha=
\begin{pmatrix}
0 & \bm \sigma \\
\bm \sigma & 0
\end{pmatrix},\qquad
\beta=
\begin{pmatrix}
1 & 0\\
0 & -1
\end{pmatrix} ,
$$
where $\bm \sigma$ are  the Pauli spin matrices.  
The semiclassical Dirac wave packet is composed of the positive energy solutions of the free Dirac equation for particles and negative energy solutions for antiparticles: 
$$
\psi_{\ssI\bm x} (\bm{p}_c,t) = \sum_\alpha \xi_{\ssI\alpha}  \psi_\ssI^\alpha  (\bm{p}_c) e^{-i\sign(q_{\ssI})p_\mu x^\mu /\hbar},
$$ 
where $p_\mu=(-E,\bm{p}_c),\ x^\mu=(t,\bm{x})$ and $\alpha=1,2.$ $I=p,a$ labels the particles ($p$) corresponding to positive energy solutions with $q_{p}=q$ and antiparticles ($a$) corresponding to negative energy solutions with  $q_{a}=-q.$ 
The coefficients  $\xi_{ \ssI\alpha} , $ are chosen to be constant. $\bm{x}_c,$ and  $\bm{p}_c,$ denote the phase space  coordinates of wave packet centre coinciding with the centre of mass. 
We define the one-form $\eta_0$ through 
$$
 \int [dx] \delta(\sign(q_{ \ssI})\bm{x}_c - \bm{x}) \Psi^\dagger_{\ssI\bm{x}}(\bm{p}_c,0) \left( -i \hbar d -\sign(q_\ssI )H^{\scriptscriptstyle{(4)}}_\ssD dt \right)\Psi_{\ssI\bm{x}}(\bm{p}_c,0) =\sum_{\alpha\beta}\xi^*_{ \ssI\alpha} \eta^{\alpha\beta}_{\ssI 0} \xi_{ \ssI,\beta}.
$$ 
$\eta_{\ssI 0}$ which is a matrix in ``spin indices"  $\alpha, \beta ,$  can be written as
\begin{equation}
\eta^{\alpha\beta}_{\ssI 0}= - \delta^{\alpha\beta}\bm{x}_c\cdot d\bm{p}_c  - \sign(q_{ \ssI})\bm A^{\alpha\beta}\cdot d{\bm p}_c -\sign(q_{ \ssI})H_{\ssD\ssI}^{\alpha\beta}dt  .
\label{et1}
\end{equation}
Here $H_{\ssD \ssI}^{\alpha\beta}$ is the projection of the Dirac Hamiltonian (\ref{hamiltonian}),  on the positive (negative) energy solutions
The matrix valued Berry gauge field is defined in terms of $\psi_\ssI $, 
\begin{equation}
\label{bgd}
\bm A^{\alpha\beta}= -i \hbar \psi_{\ssI}^{\dagger(\alpha)}(\bm p_c)\frac{\partial }{\partial {\bm p_c}} \psi_{\ssI }^{(\beta)}(\bm p_c). 
\end{equation}
It takes the same form for $I=p,a.$
By relabelling $(\bm{x}_c,\bm{p}_c)\rightarrow (\bm{x},\bm{p})$ and adding an exact differential term, the one-form (\ref{et1}) can be rewritten as
$$
\eta_{\ssI 0}=\bm p \cdot d\bm x - \sign(q_{ \ssI})\bm A \cdot d\bm p -\sign(q_{ \ssI})H_{\ssD \ssI} dt.
$$
Unless necessary the spin indices and the related unit matrix are suppressed. 
Let us 
consider the first order Hamiltonian formalism designated by the one-form
\begin{equation}
\eta_\ssI =\bm p \cdot d\bm x -\sign{(q_\ssI}) \bm A  (\bm p  ) \cdot d\bm p  +\bm a  (\bm x, \bm p , t ) \cdot d\bm  x +\phi  (\bm x, \bm p , t )dt -H_\ssI  (\bm x, \bm p , t ) dt,
\label{eta}
\end{equation} 
where $\phi,\bm a$ are electromagnetic potentials and $\sign({q_\ssI })H_{\ssI}^{\alpha \beta}\equiv H_\ssI $ denotes the projection of $H^{\scriptscriptstyle{(4)}}$  on the positive (negative) energy solutions of the free Dirac equation.
We define  the extended symplectic two-form matrix by
$$
\tilde{\omega}_{t, \ssI} = d\eta_\ssI \equiv dt \frac{\partial \eta_\ssI }{\partial t} + d \bm x \cdot \frac{\partial \eta_\ssI }{\partial \bm x }
 +  d\bm p \cdot \bm D \eta_\ssI ,
$$
where we introduced the covariant derivative
$$
\bm D \equiv \frac{\partial }{\partial \bm p}+\frac{i}{\hbar}[\bm A,\ ].
$$
By employing the one-form  (\ref{eta}), we acquire
\begin{eqnarray}
\tilde{\omega}_{t, \ssI} &=& {dp}_i \wedge{dx}_i +D_i a_j\ {dp}_i \wedge{dx}_j- \sign{(q_\ssI })G +F + \left(\frac{\partial \phi}{\partial x_i} - \frac{\partial a_i}{\partial t}\right)\  {dx}_i \wedge dt -\frac{\partial H_\ssI }{\partial x_i}\  {dx}_i \wedge dt \nonumber\\
&+& D_i \phi \ {dp}_i \wedge dt  -D_i H_\ssI \ {dp}_i \wedge dt.
\end{eqnarray}
As usual the repeated indices are summed over.
$F$ is the  curvature two-form of the  gauge field $\bm a,$ 
$$
F=\frac{1}{2}\left( \frac{\partial a_j}{\partial x_i}-\frac{\partial a_i}{\partial x_j}\right){dx}_i \wedge {dx}_j,
$$
and the Berry curvature two-form  $G=\frac{1}{2} {G_{ij}}{dp}_i \wedge {dp}_j$  is defined through the covariant derivative,  
\begin{equation}
G_{ij} =-i\hbar[D_i.D_j]=\left( \frac{\partial A_j}{\partial p_i}- \frac{\partial A_i}{\partial p_j}+\frac{i}{\hbar}[A_i,A_j]\right)= {\epsilon}_{ijk}G_k
\label{eq:G}.
\end{equation}
The matrix valued Berry gauge field is
\begin{equation}
\bm{A}=\hbar \frac{\bm{\sigma} \times \bm{p}}{2E(E+m)}, \label{bgf}
\end{equation}
The  curvature of the non-Abelian gauge field $\bm A$ is 
\begin{equation}
\bm G= \frac{\hbar m}{2E^3}\left( \bm{\sigma}+\frac{\bm{p}(\bm{\sigma}\cdot\bm{p})}{m(m+E)}\right) ,
\label{berrycurvature}
\end{equation}
which furnishes the Berry curvature via $ G_{ij} ={\epsilon}_{ijk}G_k.$ 

 Dirac Hamiltonian coupled to the external magnetic field  $\bm{B}$  and   to the  constant angular velocity of the frame $\bm \Omega$ is given as \cite{HehlNi, Bliokh},
\begin{equation*}
\label{reham}
H^{\scriptscriptstyle{(4)}} = \beta m +\bm{\alpha}\cdot\bm{ p }  -\frac{\hbar}{2}\bm{\Sigma}\cdot \bm{\Omega}-\frac{\hbar q}{2E}\bm{\Sigma}\cdot \bm{B},
\end{equation*}
where 
$\bm{\Sigma}=\begin{pmatrix} \bm{\sigma}&0\\0&\bm{\sigma}\end{pmatrix}.$
To  accomplish  the semiclassical Hamiltonian  we work in the adiabatic approximation \cite{sy,fuj} where   level crossing is absent. Then, the Dirac Hamiltonian can be diagonalized continuously at every time yielding
\begin{equation}
\label{hsm}
H_\ssI =E [1-\sign{(q_\ssI })\bm G \cdot (q\bm{B} + E\bm{\Omega}) ],
\end{equation}
Two-dimensional  unit matrix is suppressed throughout the paper. In this semiclassical approach the terms which are second or higher orders in the Planck constant are ignored.

The extended symplectic two-form $\tilde{ \omega}_{t, \ssI},$ which incorporates the dynamics of the system, lies at the heart of the Hamiltonian formalism. For the Dirac particle (antiparticle) in rigidly rotating coordinates, in the presence of the electric and magnetic fields, $\bm E,\bm B,$   it is defined as
\begin{eqnarray}
\tilde{\omega}_{t, \ssI} & = & {dp}_i \wedge {dx}_i + \frac{1}{2} \epsilon_{ijk} (q B_k + 2 {\cal E} _\ssI \Omega_k)\ {dx}_i \wedge {dx}_j -\sign{(q_\ssI })\frac{1}{2} \epsilon_{ijk} G_{k}\ {dp}_i \wedge {dp}_j \nonumber \\ 
&&+ \epsilon_{ijk} x_j\Omega_k    (\nu_\ssI )_m {dx}_i \wedge {dp}_m - (\nu_\ssI )_i\ {dp}_i \wedge dt + \frac{1}{2}  (\nu_\ssI )_i (\bm \Omega \times \bm x)^2 {dp}_i \wedge dt \nonumber\\
&&+[q\bm E+ (\bm\Omega\times \bm x)\times (q\bm B +{\cal E} _\ssI \bm\Omega)]_i\ {dx}_i\wedge dt \label{wtf},
\end{eqnarray} 
where ${\cal{E}} _\ssI=H_\ssI $ is  the dispersion relation and 
$\bm e$ denotes the effective force
\begin{equation}
\label{efel}
\bm e =  q\bm E + (\bm\Omega\times \bm x)\times (q\bm B +{\cal E} _\ssI\bm\Omega).
\end{equation}
It is composed of two parts: the Lorentz force associated to the electric field  in rotating coordinates, $\bm E^\prime =\bm E +(\bm\Omega\times \bm x)\times \bm B ,$ and the centrifugal force $ (\bm\Omega\times \bm x)\times {\cal E} _\ssI\bm\Omega.$ The ``canonical velocity"  ${\bm \nu}_\ssI$ is defined as the covariant derivative of the semiclassical Hamiltonian (\ref{hsm}):   
\begin{equation}
\bm \nu_\ssI  = \bm D H_\ssI = \frac{\bm p}{E} \left[ 1
+2\sign{(q_\ssI })\bm G \cdot \left(  q \bm B+\frac{E}{2}\bm{\Omega}\right) \right]
- \frac{ \hbar }{2E^3}   \ (q\bm B +gE\bm \Omega) \bm{\sigma} \cdot \bm{p}  . \label{numas}
\end{equation}

One can calculate the Pfaffian and  time evolutions of  phase space variables weighted with the correct measure by inspecting the Lie derivative of the volume form, which is
\begin{eqnarray}
\label{vftw}
\tilde{\Omega} _\ssI &=& \frac{1}{3!} \tilde{\omega}_{t, \ssI} \wedge \tilde{\omega}_{t, \ssI} \wedge \tilde{\omega}_{t, \ssI} \wedge dt ,
\end{eqnarray}
given in terms of extended symplectic two form, (\ref{wtf}). 
The volume form(\ref{vftw}) can be expressed as 
\begin{equation}
\label{wfpf}
\tilde{\Omega} _\ssI= (\tilde{\omega}_\ssh) _\ssI \ dV \wedge dt,
\end{equation}
where $(\tilde{\omega}_\ssh) _\ssI $  is the Pfaffian of $(6\times 6)$ matrix
\begin{equation}
\label{syma}
\begin{pmatrix}
\epsilon_{ijk} (q B_k + 2{\cal E} _\ssI \Omega_k) & -\delta_{ij}+\nu_{Ij}(\bm x \times\bm \Omega)_i \\
\delta_{ij}-\nu_{Ii}(\bm x \times\bm \Omega)_j &\ -\sign(q_\ssI )\epsilon_{ijk} G_{k}
\end{pmatrix}. 
\end{equation}
To attain the Liouville equation we need to calculate the Lie derivative of  volume form which can be carried out in two different ways. One of them is to utilize the definition of the volume form in terms of the Pfaffian, (\ref{wfpf}):
\begin{eqnarray}
{\cal{L}}_{\tilde v} \tilde{\Omega} _\ssI &=& (i_{\tilde v}  d + d i_{\tilde v} ) ((\tilde{\omega}_\ssh) _\ssI dV \wedge dt), 
\label{lievolume2}
\end{eqnarray}
where $i_{\tilde{v}}$ denotes the interior product of the vector field
\begin{equation}
\label{vf}
\tilde v= \frac{\partial}{\partial t}+\dot{\tilde {\bm x}}\frac{\partial}{\partial \bm{x}}+\dot{\tilde {\bm p}}\frac{\partial}{\partial \bm{p}}.\nonumber
\end{equation}
The other way is to employ  the  definition of  volume form  (\ref{vftw}), and directly compute its Lie derivative:
\begin{eqnarray}
{\cal{L}}_{\tilde v} \tilde{\Omega} _\ssI &=& (i_{\tilde v}  d + d i_{\tilde v} )(\frac{1}{3!} (\tilde{\omega}_t) _\ssI^3 \wedge dt)\nonumber\\
&=& \frac{1}{3!} d {\tilde{\omega}_{t, \ssI}}^3 .
\label{liouville}
\end{eqnarray}
Explicit calculation of ${\tilde{\omega}_{t, \ssI}}^3$ and the comparison of (\ref{liouville})  with  (\ref{lievolume2}), provide us   the explicit form of  Pfaffian and $ \dot{(\tilde{ \bm x}}\tilde{\omega}_\ssh) _\ssI ,$ $(\tilde{\omega}_\ssh \dot{\tilde {\bm p}}) _\ssI,$ for both particles and antiparticles which are the solutions of the equations of motion in terms of the phase space variables  $(\bm x,\bm p),$ as
\begin{eqnarray}
(\tilde{\omega}_{\ssh}) _\ssI &=&1+\sign(q_\ssI )  \ \bm{G} \cdot (q\bm{B} + 2 {\cal E} _\ssI \bm{\Omega} ) - \bm {\nu} _\ssI \cdot (\bm x \times \bm \Omega) -\sign(q_\ssI )  (  \bm {\nu} _\ssI  \cdot \bm G)(q \bm B \cdot (\bm x \times \bm \Omega)),
\label{Pfaf} \\
(\dot{\tilde{ \bm x}}\tilde{\omega}_\ssh) _\ssI&=&{\bm {\nu} _\ssI} (1 -\frac{1}{2} (\bm \Omega \times \bm x)^2) +\sign(q_\ssI )   \ \bm e \times \bm G \nonumber\\
&&+ \sign(q_\ssI )  ({\bm {\nu} _\ssI} \cdot \bm G) (q \bm B+ 2{\cal E} _\ssI \bm \Omega)(1 -\frac{1}{2} (\bm \Omega \times \bm x)^2)+ \sign(q_\ssI ) ( \bm {\nu} _\ssI \cdot \bm G) [ (\bm x \times \bm \Omega) \times \bm e ]    ,\label{msxd}\\
(\tilde{\omega}_\ssh \dot{\tilde {\bm p}}) _\ssI&=& \bm e + {\bm {\nu} _\ssI} \times (q\bm{B} + 2 {\cal E} _\ssI \bm{\Omega}) (1 -\frac{1}{2} (\bm \Omega \times \bm x)^2) \nonumber\\
&& +\sign(q_\ssI )  \bm G (\bm e \cdot (q \bm{B} + 2{\cal E} _\ssI \bm{\Omega})) - [ (\bm x \times \bm \Omega) \times \bm e ] \times {\bm {\nu} _\ssI} .  \label{mspd}
\end{eqnarray}
Equations (\ref{Pfaf})-(\ref{mspd}) will be employed in the  Boltzmann transport  equation.

\section{Boltzmann Transport Equation in Relaxation Time Approach}
\label{SCB}

To generalize the one particle dynamics of Section \ref{STDP} to many particle systems, one introduces  the distribution function $f_\ssI $ which  is defined to satisfy the Boltzmann  transport equation 
\begin{eqnarray}
\label{bte}
(\tilde{\omega}_\ssh) _\ssI \frac{\partial f_\ssI }{\partial t}+(\dot{\tilde{ \bm x}}\tilde{\omega}_\ssh) _\ssI \cdot\frac{\partial f_\ssI }{\partial \bm{x}}+(\tilde{\omega}_\ssh \dot{\tilde {\bm p}}) _\ssI\cdot\frac{\partial f_\ssI }{\partial \bm{p}}= (I_{\scriptscriptstyle{coll}}) _\ssI,
\end{eqnarray}
where $(I_{\scriptscriptstyle{coll}}) _\ssI$ denotes the collision integral. 
We adopt the relaxation time approach by choosing  the collision integral  as
\begin{equation}
(I_{\scriptscriptstyle{coll}}) _\ssI=-\frac{1}{\tau}(\tilde{\omega}_\ssh) _\ssI (f_\ssI -f^\ssI_{equ}). \label{intint}
\end{equation}
The equilibrium  distribution function $f^\ssI _{equ}$ is chosen as the Fermi-Dirac distribution 
$$f_0^ \ssI=\frac{1}{e^{[E-\sign (q_\ssI ) \mu(\bm x,t)]/kT}-1},$$
	where $\mu(\bm x,t)$ is the inhomogeneous chemical potential and the dispersion relation is approximated by ignoring the Planck constant dependence: ${\cal{E}}=E.$
	This choice of equilibrium distribution function may give the impression of being inappropriate for two reasons: $i)$ For being a Lorentz scalar $E$ should be substituted by the scalar product of momentum and velocity 4-vectors:  
	$p_\mu u^\mu \equiv E-\bm p \cdot (\bm \Omega \times \bm x) .$ However,  the symplectic two form  (\ref{wtf}) has been defined by taking into account the linear velocity  $(\bm\Omega\times\bm{x}),$  so that if one would like to keep $\bm p \cdot (\bm \Omega \times \bm x) $ term she or he should set $(\bm\Omega\times\bm{x})$  dependent terms to zero in the symplectic two form (\ref{wtf}). $ii)$ For strong magnetic field the equilibrium distribution function would possess terms which depend on magnetic field due to  quantized background, for example it may be taken as the trace of  Wigner function as  discussed recently for  similar cases in  Refs.\cite{gmss, srvw}. However,  this would mean that one permits level crossing (band mixing) \cite{g1, g2} in the definition of wave packet which is  in contradiction with the adiabatic approximation.  In the semiclassical approximation which has been  adopted the wave packet is composed of free Dirac equation solutions.

Obviously we can only approximately  solve  the kinetic equation (\ref{bte}). Actually we would like to solve it up to linear terms in $\bm e$ and the derivatives of the chemical potential.    The former  is equivalent to consider first order terms in the  electric field, $\bm E,$ and the linear velocity due to the rotation, $(\bm\Omega\times\bm{x}).$  We keep only the first order derivatives. Hence the derivatives of $\bm E^\prime$ are considered as second order although it is not  fully consistent, so that when we come to grips with the calculations of currents we will deal with the mutually parallel  magnetic field and angular velocity. 
We write  the distribution function as $f_\ssI =f^\ssI_0+f^\ssI_1, $ so that  the Boltzmann equation (\ref{bte}) turns out to be 
 \begin{eqnarray}
 \label{beeq}
 (\tilde{\omega}_\ssh) _\ssI \frac{\partial f_\ssI }{\partial t}+(\dot{\tilde{ \bm x}}\tilde{\omega}_\ssh) _\ssI \cdot\frac{\partial f_\ssI }{\partial \bm{x}}+(\tilde{\omega}_\ssh \dot{\tilde {\bm p}}) _\ssI\cdot\frac{\partial f_\ssI }{\partial \bm{p}}=-\frac{1}{\tau}(\tilde{\omega}_\ssh) _\ssI f^\ssI_1.
 \end{eqnarray}
We would like to solve for $f_1^\ssI$ by inserting the semiclassical solutions  (\ref{Pfaf})-(\ref{mspd}) into (\ref{beeq}). 
To facilitate the derivation we restrict our attention to particles  and drop the index $I.$  The solution for antiparticle distribution function is straightforward once the solution for particles is obtained. To simplify our calculations 
 we ignore the quantum corrections in (\ref{hsm}), so that we set  ${\cal{E}}=E$ and $\bm \nu=\bm p/E.$
The semiclassical Boltzmann transport equation for the Dirac particles becomes
\begin{eqnarray}
&&(1+ \bm{G} \cdot \bm{{\cal B}} ) \frac{\partial}{\partial t} (f_0+f_1)+ 
(\bm{e}+\frac{\bm{p}}{E} \times \bm{{\cal B}} + \bm{G} (\bm{e} \cdot \bm{{\cal B}} )\cdot\frac{\partial}{\partial \bm{p}} (f_0+f_1)+ \nonumber\\
&&(\frac{\bm{p}}{E}+ \bm{e} \times \bm{G} + \frac{\bm{{\cal B}}}{E} (\bm{G} \cdot \bm{p}) )\cdot \frac{\partial}{\partial \bm{x}} (f_0+f_1) =  -\frac{1}{\tau} (1+ \bm{G} \cdot \bm{{\cal B}} ) f_1,
\label{Boltzd}
\end{eqnarray} 
where we performed the relabelling $\bm{{\cal B}}\equiv q \bm{B}+2E\bm{\Omega}$.  At first sight one can think that on the right-hand side  only $f_0$ might be kept. However, in that case $\bm{{\cal B}}$ dependent terms cannot give any contribution to $f_1.$ Hence we retain the $\partial f_1/\partial \bm p$  term but treat the spatial and time derivatives of $f_1$  as  second order terms. Thus  (\ref{Boltzd}) leads to
\begin{equation}
(1+ \bm{G} \cdot \bm{{\cal B}} ) \frac{\partial f_0}{\partial t}+[\bm{e}+ \bm{G} (\bm{e} \cdot \bm{{\cal B}}) ]\cdot\frac{\partial f_0}{\partial \bm{p}} +(\frac{\bm{p}}{E} \times \bm{{\cal B}} )\cdot\frac{\partial f_1}{\partial \bm{p}} + [\frac{ \bm{p}}{E}+  \frac{\bm{{\cal B}}}{E} (\bm{G} \cdot \bm{p}) ]\cdot  \frac{\partial  f_0}{\partial \bm{x}} =  -\frac{1}{\tau} (1+ \bm{G} \cdot \bm{{\cal B}} ) f_1 .
\label{Boltzd2}
\end{equation}
We propose a solution for $f_1$ in the form
\begin{equation}
f_1= -\frac{\partial f_0}{\partial E} ( \bm{\chi} \cdot \bm{p})+\tau\frac{\partial f_0}{\partial E}\frac{\partial \mu}{\partial t},
\label{f1}
\end{equation}
where $\bm\chi$ is  linear in $\bm e$ and the gradient of $\mu$. By substituting $f_1$ with (\ref{f1}) in  (\ref{Boltzd2}), one gets
\begin{eqnarray}
-(1+ \bm{G} \cdot \bm{{\cal B}} )\frac{\partial f_0}{\partial E}\frac{\partial \mu}{\partial t}+(\bm{e} + \bm{G} (\bm{e} \cdot \bm{{\cal B}}) )\cdot\frac{\bm p}{E}\frac{\partial f_0}{\partial E}-(\frac{\bm{p} }{E}\times \bm{{\cal B}}) \cdot\bm{\chi}\frac{\partial f_0}{\partial E} + (\frac{\bm{p}}{E}+  \frac{\bm{{\cal B}}}{E} (\bm{G} \cdot\bm{p}) )\cdot  \frac{\partial\mu}{\partial \bm{x}} \frac{\partial f_0}{\partial E} \nonumber\\
=   (1+ \bm{G} \cdot \bm{{\cal B}} ) \Big(\frac{1}{\tau}\frac{\partial f^0}{\partial E} ( \bm{\chi} \cdot \bm{p})-\frac{\partial f^0}{\partial E}\frac{\partial \mu}{\partial t}\Big).
\label{Boltzd3}
\end{eqnarray}
It will be more tractable to separate $\bm \chi$  into two parts:
$$
\bm\chi= \bm \chi^0 + \bm \chi^1,
$$
where $\bm \chi^0$ is independent of $\hbar,$ and $\bm \chi^1$ is linear in  $\hbar$.  Dependence of $\bm \chi$ on the direction of the momentum vector, $\hat{\bm p},$  can only be through $\bm G,$ Therefore $\bm \chi^0 $ should be independent of $\hat{\bm p}.$  Now
by selecting terms according to the $\hbar$ order  in (\ref{Boltzd3}), we acquire the coupled equations
\begin{equation}
\bm{e}_\mu - \bm{{\cal B}} \times \bm \chi^0 =  \frac{E}{\tau} \bm \chi^0, 
\label{ki0eq}
\end{equation}
\begin{equation}
-(\frac{\bm p}{E}\times\bm{{\cal B}})\cdot\bm \chi^1+(\bm e_\mu \cdot \bm{{\cal B}})( \bm G\cdot\bm p) =\frac{1}{\tau}\bm \chi^1 \cdot\bm p+\frac{1}{\tau}( \bm{G} \cdot \bm{{\cal B}})(\bm \chi^0\cdot\bm p).
\label{ki1eq}
\end{equation}
The effective electric field (force), (\ref{efel}), and the gradient of the chemical potential behave similarly, hence we  introduced
$$
\bm e_\mu = \bm e -\bm \nabla\mu.
$$
The solution of (\ref{ki0eq}) can be shown to be \cite{anselm}
\begin{equation}
\bm \chi^0= \frac{g \bm e_\mu  - g^2 \bm{{\cal B}} \times \bm e_\mu + g^3 \bm{{\cal B}} (\bm e_\mu \cdot \bm{{\cal B}}) }{1+d^2},
\label{ki0}
\end{equation}
where $g=(\tau \slash E)$ and  $1+d^2=1+(\tau^2 \slash E^2) A^2.$
We propose $\bm  \chi^1$ to be in the form
\begin{equation}
	\bm \chi^1=g(\bm e_\mu\cdot\bm{{\cal B}})\bm G+(\bm G\cdot\bm{{\cal B}})\bm C+\bm K 
	\label{ki1sp}
\end{equation}
and then  solve (\ref{ki1eq}) for $\bm C$ and $\bm K.$  They  are necessarily independent of $\hat{\bm p}$, otherwise  $\bm \chi^1$ would  not  satisfy the equality
$$
(\bm p \times \bm{{\cal B}})\cdot \frac{\partial (\bm p \cdot \bm \chi^1)}{\partial \bm p} = (\bm p \times \bm{{\cal B}})\cdot \bm \chi^1,
$$
which has been assumed in obtaining (\ref{Boltzd3}).
Plugging (\ref{ki1sp}) into (\ref{ki1eq}) yields the equation
\begin{eqnarray}
	-g(\bm e_\mu\cdot\bm{{\cal B}})(\frac{\bm p}{E}\times\bm{{\cal B}})\cdot\bm G-(\bm G\cdot\bm{{\cal B}})(\frac{\bm p}{E}\times\bm{{\cal B}})\cdot\bm C-(\frac{\bm p}{E}\times\bm{{\cal B}})\cdot\bm K&=& \frac{1}{\tau}(\bm G\cdot\bm{{\cal B}})(\bm C\cdot\bm p)+\frac{1}{\tau}(\bm K\cdot\bm p)  \nonumber \\
&& +\frac{1}{\tau}(\bm G\cdot\bm{{\cal B}})(\bm \chi^0\cdot\bm p).
	\label{CK}
\end{eqnarray}
We observe that the terms containing the vector $\bm C$ can be written as
\begin{equation}
\nonumber
	-\frac{E}{\tau}\bm \chi^0-\bm{{\cal B}}\times\bm C=\frac{E}{\tau}\bm C,
\end{equation} 
which  is in the same form with equation (\ref{ki0eq}). Hence, one can attain  $\bm C$ as
\begin{equation}
	\bm C=\frac{1}{1+d^2}(-\bm \chi^0+g\bm{{\cal B}}\times\bm \chi^0-g^2(\bm{{\cal B}}\cdot\bm \chi^0)\bm{{\cal B}}) .
    \label{C}
\end{equation}
Having obtained $\bm C$ in terms of $\bm \chi_0,$ we now examine  the terms related to $\bm K$ in (\ref{CK}). The equation for $\bm K$ can be written as
\begin{equation}
	\label{eqK}
	-g\frac{\bm p}{E}\cdot(\bm{{\cal B}}\times\bm G)(\bm e_\mu\cdot\bm{{\cal B}})-\frac{\bm p}{E}\cdot(\bm{{\cal B}}\times\bm K)=\frac{1}{\tau}\bm K\cdot\bm p.
\end{equation}
To solve it  by imitating the solution of (\ref{ki0eq}), we must make sure that  $\bm K$ is independent of $\hat{\bm p}.$ In fact, by inspecting the  first term of (\ref{eqK}) which is  proportional to $\bm{G},$  one can observe that there is a vanishing part,  so that (\ref{eqK})  reduces to
\begin{equation}
\nonumber
-g\frac{m\hbar}{2E^3}(\bm{{\cal B}}\times\bm\sigma)(\bm e_\mu\cdot\bm{{\cal B}})-(\bm{{\cal B}}\times\bm K)=\frac{E}{\tau}\bm K.
\end{equation}
Thus, we solve (\ref{eqK}) for $\bm K$ as
\begin{equation}
\bm K=-\frac{m\hbar}{2E^3}\frac{g^2(\bm{{\cal B}}\times\bm\sigma)(\bm e_\mu\cdot\bm{{\cal B}})-g^3(\bm{{\cal B}}\times(\bm{{\cal B}}\times\bm{\sigma}))(\bm e_\mu\cdot\bm{{\cal B}})}{1+d^2}.
\label{K}
\end{equation}
Plugging (\ref{C}) and (\ref{K}) into (\ref{ki1sp}), one obtains $\bm \chi_1$.  Then by employing 
 (\ref{ki0}) and (\ref{ki1sp}) in (\ref{f1})  we establish the first order distribution function in the relaxation time approach as
\begin{eqnarray}
f_1=&& -\frac{\partial f_0}{\partial E}\Bigg\{g(\bm{{\cal B}}\cdot\bm e_\mu)(\bm G\cdot\bm p)-\tau \frac{\partial \mu}{\partial t} \nonumber \\
&&+\frac{\bm{p}\cdot\Big(g\bm e_\mu-g^2\bm{{\cal B}}\times\bm e_\mu+g^3(1-\bm G\cdot\bm{{\cal B}})(\bm{{\cal B}}\cdot\bm e_\mu)\bm{{\cal B}}-g^2\frac{m\hbar}{2E^3} (\bm{{\cal B}}\cdot\bm e_\mu)[(\bm{{\cal B}} \times\bm \sigma)-g\bm{{\cal B}}\times(\bm{{\cal B}} \times\bm \sigma)]\Big)}{1+d^2} \nonumber \\
&&-\frac{(\bm{G}\cdot\bm{{\cal B}})}{(1+d^2)^2}\Big(g(1-g^2A^2)\bm e_\mu\cdot\bm{p}-2g^2(\bm{{\cal B}}\times\bm e_\mu)\cdot \bm{p}+2g^3(\bm{{\cal B}}\cdot\bm e_\mu)(\bm{{\cal B}}\cdot \bm{p})\Big) \Bigg\}\nonumber .
\end{eqnarray}
As far as  the  currents linear in $\bm e_\mu$ are concerned,  it is sufficient to deal with $f=f_0+f_1$ for particles and correspondingly for antiparticles. 

\section{Spin and Spin  Current Densities}
\label{SCD}
Equipped with the solution of the Boltzmann equation the particle and antiparticle number densities are defined as
\begin{equation}
n_\ssI  =\int \frac{d^3p}{(2\pi\hbar)^3}\Tr[(\tilde{\omega}_{1/2})_\ssI f_\ssI ]. \label{pnccf}
\end{equation}
Total number density is
$$n=\sum_\ssI \sign(q_\ssI )n_\ssI .$$
Collisions should conserve the number density. Therefore,  due to our choice of   collision integral (\ref{intint}), we need to  constrain the nonequilibrium distribution functions to satisfy 
\begin{equation}
\sum_\ssI \sign(q_\ssI ) \int \frac{d^3p}{(2\pi\hbar)^3}\Tr[(\tilde{\omega}_{1/2})_\ssI f^\ssI _1]=0. 
\label{consteq}
\end{equation}
Then the particle (antiparticle) number density (\ref{pnccf}) involves only  the equilibrium distribution function:
\begin{equation}
n_\ssI  =\int \frac{d^3p}{(2\pi\hbar)^3}\Tr[(\tilde{\omega}_{1/2})_\ssI f^\ssI _0]. \label{pncc}
\end{equation}
The particle (antiparticle) current density is  similarly defined:
\begin{equation}
\bm{j} _\ssI =\int \frac{d^3p}{(2\pi\hbar)^3}\Tr[(\dot{\tilde{ \bm x}}\tilde{\omega}_\ssh) _\ssI f_\ssI ]. \nonumber
\end{equation}
Total current density is
 $$\bm{j}=\sum_\ssI \sign(q_\ssI )\bm{j} _\ssI.$$ 
One can show that total number and current densities  satisfy the continuity equation
$$
\frac{\partial n}{\partial t }+\bm \nabla \cdot {\bm j}=0,
$$
on account of the consistency  condition (\ref{consteq}) and letting $\bm B,\bm \Omega$   be mutually parallel.

The $4\times4$ spin matrices of  the  Dirac particles  are
  $\frac{\hbar}{2}\bm\Sigma .$
In order to attain the semiclassical formalism we have projected all of the $4\times4$ matrix-valued physical quantities
on the positive (negative) energy solutions of the Dirac equation. Hence, 
it is appropriate to define the spin matrix in the semiclassical formalism as the projection of $\frac{\hbar}{2}\bm\Sigma$ onto the positive (negative) energy solutions which yields the Pauli matrices, $\frac{\hbar}{2}\bm\sigma .$ 
We would like to study  spin currents. To this aim we first define  the spin density  of particles and antiparticles having spin $\hbar/2$ in the  $a$-direction, where $a=x,y,z,$ as
\begin{eqnarray}
\label{snd}
n^a_\ssI  =\frac{\hbar}{2}\int \frac{d^3p}{(2\pi\hbar)^3}\Tr[(\sigma_a\tilde{\omega}_{1/2})_\ssI f^\ssI _0].
\end{eqnarray}
The total spin density is  $$n^a=\sum_\ssI \sign(q_\ssI )n^a_\ssI .$$
Observe that it is defined only in terms of $f^\ssI _0$ as the particle (antiparticle) number density (\ref{pncc}). Thus  for consistency we should demand that 
\begin{equation}
\sum_\ssI \sign(q_\ssI )\int \frac{d^3p}{(2\pi\hbar)^3}\Tr[\sigma_a(\tilde{\omega}_{1/2})_\ssI f^\ssI _1]=0.
\label{consteq1}
\end{equation}
Actually, this condition  yields the time evolution of the inhomogeneous chemical potential, $\mu (\bm x ,t).$ 
Naturally, we  define the spin current densities in the $a$-direction  as
\begin{equation}
\label{totj}
\bm j^a_\ssI =\frac{\hbar}{2}\int \frac{d^3p}{(2\pi\hbar)^3}\Tr[\sigma_a(\dot{\tilde{ \bm x}}\tilde{\omega}_\ssh) _\ssI f_\ssI ].
\end{equation}
It is convenient to separate the spin currents into two parts depending on $f^\ssI _0$ and $f^\ssI _1$ as
\begin{eqnarray}
\bm{J^a} _\ssI&=&\frac{\hbar}{2}\int \frac{d^3p}{(2\pi\hbar)^3}\Tr[\sigma_a(\dot{\tilde{ \bm x}}\tilde{\omega}_\ssh) _\ssI f^\ssI _0],  \label{ecc}\\
{\bm {{\cal{ J}}}}^{a} _\ssI&=&\frac{\hbar}{2}\int \frac{d^3p}{(2\pi\hbar)^3}\Tr[\sigma_a(\dot{\tilde{ \bm x}}\tilde{\omega}_\ssh) _\ssI f^\ssI _1].  \label{necc}
\end{eqnarray}
Thus  $\bm j^a_\ssI =\bm J^a_\ssI  +\bm {{\cal{ J}}}^a_\ssI $  give the spin current densities for particles and antiparticles.
The total spin current density is $$\bm j^a=\sum_\ssI \sign(q_\ssI )\bm j^a_\ssI .$$
At $T=0$ there is no contribution from antiparticles to the number and current densities  as well as to the spin and spin current  densities,  only particles contribute to integrals.  We will consider two-dimensional and three-dimensional conductors at $T=0.$ Therefore, in the sequel we will drop the index $I$ and deal only with particle contributions to spin and spin current densities.

\section{Spin Currents in Two-Dimensional Conductors}
\label{2D}

Two-dimensional condensed matter systems play an important role in realizing the spin currents. The effects of rotation alongside with the external magnetic fields in such systems have been a focus of attention. In  Refs.\cite{mism, mism2} a model to realize spin currents by mechanical  rotation in  Pt-films was proposed and  to investigate the role of mechanical rotation in generation of spin currents, a rotating disk with the Pt-film attached to it has been considered. The external magnetic field is taken along the rotation axis. This model predicts spin currents in the radial and azimuthal directions. We would like to display the application of our formalism to the Dirac particles at $T=0$ confined to move in two-dimensions inspired by such experimentally realizable condensed matter systems and find out the consequences of rotation and  external electromagnetic fields.

Our formalism is essentially three-dimensional, hence the definitions of the spin number  and current densities, (\ref{snd}), (\ref{totj}), are given in terms of the  three-dimensional momentum space integrals. Nevertheless, one can easily adapt these definitions to  two-dimensional systems by setting the third component of momentum  to zero, $p_z=0,$ then integrating over the remaining  momentum components and confining the geometry to two-dimensions by taking $z=0$.  We restrict our attention to the circular geometry: $\bm{x}=R\bm{\hat \rho},$ where  $(\rho,\phi)$ denote the polar coordinates and $R$ is constant. 
Moreover, magnetic field and angular velocity are chosen to be perpendicular to 
this circular plane. 
Under these conditions $\bm{e}=q\bm{E}+(E\Omega^2R+q\Omega BR)\bm{\hat \rho}$.To simplify our discussion let the external electric field, $\bm{E},$ lie in the $xy$-plane. Note that in Ref.\cite{mism2} spin currents are calculated for $\bm E=0.$ Both $\bm B$ and $\bm \Omega$ point in the positive $z$-direction, thus it is natural to deal with the third component of spin, $\frac{\hbar}{2}\sigma_z.$
The two-dimensional  spin  and spin current densities 
are denoted $\tilde{n}^z$  and  ${\tilde{\bm j}}^z=\tilde{ \bm J^z} +\tilde{\bm {{\cal{ J}}}}^z.$

We defined spin density in terms of  equilibrium distribution function, (\ref{snd}), so that the number of spin-up particles should not be altered by collisions, (\ref{consteq1}). Thus we demand that the following condition is satisfied,
\begin{eqnarray}
\int\frac{d^2p}{(2\pi\hbar)^2}Tr[ \sigma_z \tilde{\omega}_{1/2}f_1]_{z=p_z=0}&=&
\int\frac{d^2p}{(2\pi\hbar)^2}Tr[ \sigma_z(\bm G\cdot\bm{{\cal B}})\tau \frac{\partial f_0}{\partial E}\frac{\partial \mu}{\partial t}] \nonumber \\
&=&
\frac{\tau}{2\pi\hbar}\frac{m}{\mu^2}(qB+2\mu\Omega)\frac{\partial \mu}{\partial t}=0. \nonumber
\end{eqnarray}
Therefore, $\mu$ is independent of time at the first order. The spin  density is readily calculated as
\begin{eqnarray}
	\tilde n^z= \frac{\hbar}{2}\int \frac{d^2p}{(2\pi\hbar)^2}Tr[\sigma_z\tilde{\omega}_{1/2}f_0]&=&\frac{m}{4\pi}qB\int_{m}^{\mu}\frac{dE}{E^2}+\frac{2\Omega m}{4\pi}\int_{m}^{\mu} \frac{dE}{E} \nonumber\\
	&=&B\frac{q}{4\pi}(\frac{\mu-m}{\mu})+\Omega\frac{m}{2\pi}\ln \frac{\mu}{m}. \nonumber
\end{eqnarray}
Observe that it depends  linearly on   magnetic field and  angular velocity. The spin current  density arising from the equilibrium distribution function is 
\begin{eqnarray}
\tilde {\bm J}^z&=&\frac{\hbar}{2}\int \frac{d^2p}{(2\pi\hbar)^2}Tr[\sigma_z(\bm{e}\times\bm{G})]f_0 \nonumber\\
&=&\frac{q}{4\pi}(\frac{\mu-m}{\mu}) \bm{E} \times \bm{\hat z}-\frac{q\Omega BR}{4\pi}(\frac{\mu-m}{\mu})\bm{\hat\phi}\label{jf02}-\frac{\Omega^2Rm}{4\pi}\ln(\frac{\mu}{m})\bm{\hat\phi}.
\label{jf02d}
\end{eqnarray}
The first term is the spin Hall current in an inertial  reference frame  which has been obtained in Ref. \cite{ccn} in the nonrelativistic limit. For the rest of this section  let the electric field possess only radial component: $\bm E =E_\rho \hat{\bm \rho}.$  Now we can unify the first and the second terms of  (\ref{jf02d}) as 
$ \tilde {\bm J}^z_{\ssS \ssH} =\sigma_{\ssS\ssH} \hat{\bm z}\times \bm E^\prime,$ where  
\begin{equation}
\label{elp}
\bm E'= (E_\rho+\Omega BR ) \hat{\bm \rho}
\end{equation}
 is the electric field in rotating coordinates and
 \begin{equation}
 \label{shc}
 \sigma_{\ssS \ssH} =-\frac{q}{4\pi}\frac{\mu-m}{\mu}
 \end{equation}
is the spin Hall conductivity.  At zero temperature the chemical potential can be written in terms of the Fermi momentum $k_\ssF $
 as $\mu=\sqrt{k^2_\ssF+m^2}.$  
Hence when we  consider the nonrelativistic limit the spin Hall conductivity yields $\sigma_{\ssS \ssH}\approx -\frac{q\hbar}{4 m^2}\tilde{n},$  where $\tilde{n}$ is the particle number density in two dimensions. 
A similar  result is obtained in \cite{mism, mism2} in the nonrelativistic limit for a vanishing electric field.
In the limit $m \ll \mu$ one obtains $\sigma_{\scriptscriptstyle{SH}}=-\frac{q}{4\pi}.$ As we discussed in Ref.\cite{om-elif}, it is given by  a topological invariant called  the spin Chern number \cite{prodan,ezawa} which can be expressed as the sum of two  first Chern numbers  (see \cite{biz-ann} and the references therein).
The last term in (\ref{jf02d}) is obviously associated to the fictitious centrifugal force.

Next, we turn back to the calculation of the spin current density. 
We would like to emphasize the fact that the spin dependence of the kinetic equation is due to the Berry curvature. Hence the terms which contribute to spin currents are the ones which involve the Berry curvature. They may be present either in the velocity $\dot{\tilde{ \bm x}}\tilde{\omega}_\ssh$ or in $f_1.$

 The contributions arising from the  collisions will include various orders of the relaxation time, $\tau,$ as can be seen by examining  (\ref{f1}). Recall that the direction of the electric field is chosen in the radial direction, so that the electric field  in the rotating frame is given by (\ref{elp}). Then one can  calculate ${\tilde{\bm {{\cal{J}} }}}^{z}$  as
\begin{eqnarray}
{\tilde{\bm {{\cal{J}} }}}^{z}
=&&\frac{\hbar}{2}\int \frac{d^2p}{(2\pi\hbar)^2}\frac{\hbar m}{E^4}\frac{(qB+2E\Omega)}{(1+d^2)^2}\frac{\partial f_0}{\partial E}\bm{p}\Big(g(1-g^2{{\cal B}}^2)\bm e_\mu\cdot\bm{p}-2g^2(\bm{{\cal B}}\times\bm e_\mu)\cdot \bm{p}\Big) \nonumber\\
=&&\frac{m}{8\pi}\int dE \frac{\partial f_0}{\partial E} \frac{E^2-m^2}{E^3}\frac{g(qB+2E\Omega)}{(1+d^2)^2}
 \Big[\left(qE^\prime+ER\Omega^2\right)
\left\{
(1-g^2{{\cal B}}^2) \bm{\hat\rho}+2g (qB+2E\Omega) \bm{\hat\phi}\right\}\nonumber \\
          &&-\left\{(1-g^2{{\cal B}}^2)\bm{\nabla}\mu+2g(qB+2E\Omega)\bm{\hat{z}}\times\bm{\nabla}\mu\right\}\Big]\nonumber\\
=&&\frac{m(\mu^2-m^2)}{8\pi\mu^3}
\frac{g_\mu {{\cal B}}_\mu}{(1+d_\mu^2)^2}  
\Big[ (1-g_\mu^2{{\cal B}}_\mu^2) \left( q\bm E^\prime-\bm{\nabla}\mu+\mu R\Omega^2 \bm{\hat\rho}\right)\nonumber\\
 &&+2g_\mu {{\cal B}}_\mu  \left(\bm{\hat{z}}\times (q\bm E^\prime-\bm{\nabla}\mu) +\mu R\Omega^2 \bm{\hat\phi} \right) \Big]
 \label{jf12}
\end{eqnarray}
where we introduced the short-handed notations ${{\cal B}}_\mu=(qB+2\mu\Omega),$ $g_\mu=\frac{\tau}{\mu}$ and $d_\mu=g_\mu {{\cal B}}_\mu .$ Equation (\ref{jf12}) involves  terms both in the radial and azimuthal direction. Actually, there are  terms both parallel and perpendicular to $\bm E^\prime,$ in contrast to (\ref{jf02d}) which has only a perpendicular term.  
Let us investigate further the terms which are linear in $\bm E^\prime,$ by first decomposing them into their $x$- and $y$-components as
\begin{eqnarray}
	 (\tilde{\cal{J}}_{\ssE^\ssprime})_x^{z}&=&a_1E_x^\prime-a_2E_y^\prime \nonumber\\
	(\tilde{\cal{J}}_{\ssE^\ssprime})_y^{z}&=&a_1E_y^\prime+ a_2E_x^\prime \nonumber
\end{eqnarray}
where 
\begin{equation}
\label{a1a2}
	a_1=\frac{mq\hbar}{8\pi}\frac{\mu^2-m^2}{\mu^3}  \frac{g_\mu {{\cal B}}_\mu (1-g_\mu^2{{\cal B}}_\mu^2)}{(1+d_\mu^2)^2},\ \ \  a_2=\frac{2mq\hbar}{8\pi}\frac{\mu^2-m^2}{\mu^3}\frac{{{\cal B}}^2_\mu g^2_\mu}{(1+d_\mu^2)^2} .
\end{equation}
To study the Hall-like current densities we set $(\tilde{\cal{J}}_{\ssE^\ssprime})_y^{z}=0$ and express $E^\prime_x=-\frac{a_1}{a_2}E_y^\prime $. Obviously, we suppose that the $\bm E^\prime,$ term in the current density generated by the equilibrium distribution function (\ref{jf02d}) is suppressed. Then, the  current density in the $x$-direction is given by the electric field $E_y^\prime$   as
\begin{equation}
 (\tilde{\cal{J}}_{\ssE^\ssprime})_x^{z}= -\left(\frac{a_1^2}{a_2}+a_2\right)E^\prime_y\equiv-\sigma_{\ssS\ssH}^{(1)} E^\prime_y .
\label{JEperp}
\end{equation}
The spin Hall conductivity  can be shown to be
\begin{equation}
\sigma^{(1)}_{\scriptscriptstyle{SH}}=\frac{qm}{16\pi}\frac{\mu^2-m^2}{\mu^3}.
\label{shc1}
\end{equation}
It is still independent of the fields and can be expressed in terms of the particle number density as $\sigma^{(1)}_{\scriptscriptstyle{SH}}=\frac{q\hbar m}{8\mu^3}\tilde{n}$. However for large $\mu$ it approaches to zero.
To obtain the current density   parallel to  the electric field we set $(\tilde{\cal{J}}_{\ssE^\ssprime})_y^{z}=0$  and  express $E^\prime_y= -\frac{a_2}{a_1}E^\prime_x, $ so that
$$ 
(\tilde{\cal{J}}_{\ssE^\ssprime})_x^{z}= \frac{qm}{8\pi}\frac{\mu^2-m^2}{\mu^3}\frac{g_\mu {{\cal B}}_\mu}{1-g_\mu^2{{\cal B}}_\mu^2}E^\prime_x.$$
This is an analogue of the  Ohm's law  for the spin current in rotating coordinates and in external magnetic field. 

The effects caused by  the centrifugal force dependent term in (\ref{jf12}) can be studied  by substituting $\bm E^\prime$  with $\mu R\Omega^2 \bm{\hat \rho,}$ in the  discussions given above. Also, the gradient of the chemical potential behaves similar to $\bm E^\prime$ as can be seen by inspecting (\ref{jf12}).

The continuity equation  can be shown to yield 
\begin{eqnarray}
\frac{\partial \tilde n^z}{\partial t} + \bm{\nabla}\cdot\tilde{\bm j^z}= 0. \nonumber
\end{eqnarray}
Therefore, the spin current is conserved at the first order.

\section{Spin Currents in Three-Dimensional Conductors}
\label{3D}

Although it is quite relevant to deal with spin currents generated by rotations in two dimensional conductors,  in some cases it would be useful to deal with the three dimensional ones.  Therefore, in this section we would like to focus on  the  spin  and spin current densities generated  in the three dimensional  conductors at $T=0$. To keep the discussion general let us deal with the spin matrix $\frac{\hbar}{2}\sigma_a$ in an arbitrary direction $\bm{\hat a}=\left(\hat{\bm x},\hat{\bm y}, \hat{\bm z} \right).$ By integrating over the angular part
the spin  density (\ref{snd})  which is given by the equilibrium distribution function, $f_0,$   leads to
\begin{eqnarray}
n^a=&& \frac{\hbar}{2}\int \frac{d^3p}{(2\pi\hbar)^3}Tr[\sigma_a(1+\bm{G}\cdot(q\bm B+ 2E\bm\Omega))f_0]\nonumber\\
=&&\frac{1}{4\pi^2\hbar}\int p^2dp\Big( \frac{m}{E^3}+\frac{p^2}{3E^3(E+m)}\Big)(qB_a+2E\Omega_a)f_0 .\nonumber
\end{eqnarray}
To acquire a nonvanishing spin current in a certain direction
the external magnetic field $\bm{B}$ or  the angular velocity $\bm \Omega$ should possess a nonvanishing  component  along  the direction of  the spin. Actually, we deal only with the mutually parallel  $\bm{B}$ and  $\bm \Omega.$   Now by performing the rest of the integrals at $T=0,$ one can show that the spin density is
\begin{eqnarray}
n^a&=&\frac{qmB_a}{12\pi^2\hbar}\Big(\frac{(\mu-2m) \sqrt{\mu^2-m^2}}{m}+2m\ln(\frac{\sqrt{\mu^2-m^2}+\mu}{m})+m(\tan^{-1}(\frac{m}{\sqrt{\mu^2-m^2}})-\frac{\pi}{2}) \Big)\nonumber\\
&&+\frac{m^2\Omega_a}{12\pi^2\hbar}\Big(\frac{(4m+\mu)\sqrt{\mu^2-m^2}}{m^2}-\ln(\frac{\sqrt{\mu^2-m^2}+\mu}{m})+4(\tan^{-1}(\frac{m}{\sqrt{\mu^2-m^2}})-\frac{\pi}{2})\Big)
\label{3dna}
\end{eqnarray}
The structure of the magnetic field and angular velocity dependent terms  are quite  similar.

To obtain the spin current densities we  insert the semiclassical velocity (\ref{msxd}) into (\ref{ecc}) and (\ref{necc}) with the appropriate distribution functions. Actually, there is  only one term which may give  contribution  to the equilibrium current density (\ref{ecc}):
\begin{eqnarray}
\bm J^a&= &\frac{\hbar}{2}\int \frac{d^3p}{(2\pi\hbar)^3}Tr[\sigma_a(\bm{e}\times\bm{G})]f_0 \nonumber\\
&=&\frac{q}{12\pi^2\hbar}[(\bm E+(\bm\Omega\times\bm x)\times\bm B)\times\bm{\hat a}]\Big(\sqrt{\mu^2-m^2}(1-\frac{2m}{\mu})+2m\ln(\frac{\sqrt{\mu^2-m^2}+\mu}{m})\nonumber\\
&&+m(\tan^{-1}(\frac{m}{\sqrt{\mu^2-m^2}})-\frac{\pi}{2}) \Big)\nonumber\\
&&+\frac{1}{12\pi^2\hbar}[(\bm\Omega\times\bm x)\times \bm \Omega)\times\bm{\hat a}]\Big((4m+\mu)\sqrt{\mu^2-m^2}-m^2\ln(\frac{\sqrt{\mu^2-m^2}+\mu}{m})\nonumber \\
&&+4m^2(\tan^{-1}(\frac{m}{\sqrt{\mu^2-m^2}})-\frac{\pi}{2})\Big)
\label{3djaf0}
\end{eqnarray}
The first term is  perpendicular to the electric field in rotating coordinates and the second one is  perpendicular  to  the fictitious centrifugal force. Obviously they both  are perpendicular to the  spin direction which is considered.

Before proceed to calculate the current density arising from $f^1,$ let us examine  the consistency condition (\ref{consteq1}) for $f^1.$ One can easily observe that  after integrating over the momentum space and  keeping the first order terms, there remains only one  term  
\begin{eqnarray}
\int \frac{d^3p}{(2\pi\hbar)^3}Tr[\sigma_a\tilde{\omega}_{1/2}f_1]&=&\int\frac{d^3p}{(2\pi\hbar)^3}Tr[ \sigma_a(\bm G\cdot\bm{{\cal B}})f_1] \nonumber\\
&=&\frac{\tau}{6\pi^2\hbar^2}\frac{2m+\mu}{\mu^2}\sqrt{\mu^2-m^2}(qB_a+2\mu\Omega_a)\frac{\partial \mu}{\partial t}=0,
\label{consteq3d}
\end{eqnarray}
which states that $\mu$ is independent of time.
 Once we integrate over the angular variables, the spin current density (\ref{necc}) which is  linear in $\bm e_\mu,$   turns out to be
\begin{eqnarray}
\bm{{\cal{ J}}}^{a}&= &\frac{\hbar}{2} \int \frac{d^3p}{(2\pi\hbar)^3}Tr[\sigma_a(\frac{\bm p}{E}+\frac{\bm{{\cal B}}}{E}(\bm{G\cdot p}))f_1] \nonumber\\
&=&\frac{1}{12\pi^2\hbar}\int dp \frac{p^4}{E^4}\frac{\partial f_0}{\partial E}\Bigg[\frac{\bm{{\cal B}}\cdot\bm e_\mu}{(1+d^2)}g^2\Big(\frac{2g}{5}(E-m){{\cal B}}_a\bm{{\cal B}}-mg{{\cal B}}^2\bm{\hat a}+\bm{{\cal B}}\times\bm{\hat a}\Big)-(\bm{{\cal B}}\cdot\bm e_\mu)g\frac{m}{5}\bm{\hat a}\nonumber\\
&&-\frac{2Eg}{(1+d^2)}\Big({e_{\mu}}_a-g(\bm{{\cal B}}\times\bm e_\mu)_a+g^2(\bm{{\cal B}}\cdot\bm e_\mu){{\cal B}}_a\Big)\bm{{\cal B}}\nonumber\\
&&+\frac{1}{(1+d^2)^2}\frac{g}{5}\Big((E-4m)\Big[(1-g^2{{\cal B}}^2){{\cal B}}_a\bm {e_{\mu}}-2g{{\cal B}}_a(\bm{{\cal B}}\times\bm e_\mu)+2g^2(\bm{{\cal B}}\cdot\bm e_\mu){{\cal B}}_a\bm{{\cal B}}\Big]\nonumber\\
&&+(E-m)\Big[(1-g^2{{\cal B}}^2)e_{\mu a}\bm{{\cal B}}-2g(\bm{{\cal B}}\times\bm e_\mu)_a\bm{{\cal B}}+2g^2(\bm{{\cal B}}\cdot\bm e_\mu){{\cal B}}_a\bm{{\cal B}}\Big]\Big)\Bigg] .
\label{3djaf1}
\end{eqnarray}
The spin  density (\ref{3dna}) and the spin current density given as  the sum of (\ref{3djaf0}),(\ref{3djaf1}), satisfy the continuity equation  
$$
\frac{\partial  n^a}{\partial t} + \bm{\nabla}\cdot \bm j^a=0.
$$
Hence the spin is conserved at the first order.

To render the discussion of this linear  spin current density  comprehensible, let us deal with the spin in the \mbox{$z$-direction,}  by choosing   $\bm{{\cal B}}$ to have a nonvanishing component only in  the same direction: $\bm{B}=B\bm{\hat z}$ and  $\bm{\Omega}=\Omega\bm{\hat z}.$  Under these conditions the complicated expression (\ref{3djaf1}) for the spin current arising from the collisions,  becomes more accessible when we further impose conditions on the effective electric force $\bm e_\mu$. To this aim, we first take $\bm e_\mu$  in the same direction with $\bm B$ and $\bm \Omega$:
\begin{equation}
	{\cal{ J}}^{z}_{3}
	=\frac{\tau}{60\pi^2\hbar}\frac{(\mu^2-m^2)^{3/2}}{\mu^4}\Big[3(\mu+2m)+\frac{2m\tau^2 (qB+2\mu\Omega)^2}{\mu^2+\tau^2(qB+2\mu\Omega)^2}\Big] (qB+2\mu\Omega)\left(qE_3-\frac{\partial \mu}{\partial x_3}\right) . \nonumber
\end{equation}
It results in a current in the third-direction, as long as  the component of electric field or the gradient of the chemical potential in the third-direction  exists.
The other case which is instructive to explore is when   $\bm e_\mu$ is perpendicular to $\bm B$ and $\bm \Omega$: 
\begin{eqnarray}
{\cal{ J}}^{z}_{i}
=&&\frac{\tau}{60\pi^2\hbar}\frac{ (4m-\mu)(\mu^2-m^2)^{3/2}}{(\mu^2+\tau^2(qB+2\mu\Omega)^2)^2 }(qB+2\mu\Omega)\Big[(1-\frac{\tau^2}{\mu^2}(qB+2\mu\Omega)^2)\\
&&\times(q\bm E^\prime 
+\mu(\bm{\Omega}\times\bm {x})\times\bm\Omega-\bm{\nabla}\mu )_{i}-\frac{2\tau}{\mu}(qB+2\mu\Omega) (\bm{\hat{z}}\times (q\bm E^\prime+\mu(\bm{\Omega}\times\bm {x})\times\bm\Omega-\bm{\nabla}\mu ))_{i}\Big]\nonumber ,
\label{j3perp}
\end{eqnarray}
where  $i=1,2.$ Similar to the two-dimensional case in Section \ref{2D}, we can decompose the electric field part of (\ref{j3perp}) into its $x$ and $y$ components as 
$ ({\cal{J}}_{\ssE^\ssprime})_x^{z}=b_1E_x^\prime-b_2E_y^\prime, \ 
({\cal{J}}_{\ssE^\ssprime})_y^{z}=b_1E_y^\prime+ b_2E_x^\prime $
where
\begin{eqnarray}
 b_1&=&\frac{ (4m-\mu)(\mu^2-m^2)^{3/2}}{60\pi^2\hbar }\frac{\tau(qB+2\mu\Omega)}{(\mu^2+\tau^2(qB+2\mu\Omega)^2)^2}(1-\frac{\tau^2}{\mu^2}(qB+2\mu\Omega)^2),\nonumber\\
 b_2&=&\frac{(4m-\mu)(\mu^2-m^2)^{3/2}}{60\pi^2\hbar }\frac{(qB+2\mu\Omega)^2}{(\mu^2+\tau^2(qB+2\mu\Omega)^2)^2}\frac{2\tau^2}{\mu}\nonumber .
\end{eqnarray}
Then the  spin Hall-like conductivity can be read from (\ref{JEperp})  as
$$\sigma^{\perp}_{\ssS\ssH}= \frac{(4m-\mu)}{120\pi^2\hbar}\frac{(\mu^2-m^2)^{3/2}}{\mu^3}.$$
It is independent of $\bm B$ and $\bm{\Omega}$. The gradient of the chemical potential behaves similar to the electric field $\bm E^\prime$.

\section{Conclusions}
\label{CONC}

The spin and spin current densities  of the Dirac particles in the presence of external electromagnetic fields, in rotating coordinates are studied. The effects caused by the particles which rotate because they have been subjected to mechanical rotations or due to vorticity when they show fluid behaviour, can equivalently be viewed as effects arising due to  observation of the particles in a rotating frame. 
An intuitive  approach to  study the spin dependent phenomena  of the Dirac particles in rotating frames is offered by the semiclassical kinetic theory proposed in Ref.\cite{oee}.  In this formalism  the velocities of phase space variables are matrix-valued in spin indices, so that they lead to a  matrix valued Boltzmann transport equation. We studied this kinetic equation within  the relaxation time method. We obtained the matrix-valued distribution functions  up to terms linear in the electric field in rotating coordinates and the derivatives of the chemical potential.  The spin and spin current densities are established for two and there dimensional conductors. They yield similar effects. We showed that in two-dimensional media the equilibrium distribution functions gives rise to the spin Hall effect associated to the electric fields in rotating coordinates. The relaxation time dependent terms of spin current densities vanish when one switches off the external magnetic field and  rotations.  These terms which are at least linear in the magnetic field $\bm B$ and the angular velocity $\bm \Omega,$  yield the Hall-like current whose conductivity is independent of them. These conductivities depend only on the mass of the Dirac particles and the chemical potential.  However,  in contrast to the ordinary spin Hall conductivity, it is not a topological invariant in the $\mu \gg m$ limit. At zero temperature the chemical potential can be written in terms of the Fermi momentum $k_\ssF$ as $\mu=\sqrt{k^2_\ssF+m^2}.$ We use the Fermi wavenumber value of  $k_F=10^{10}  {\rm m}^{-1}$ for Pt \cite{tm} to estimate spin Hall conductivity and spin currents. Equation (\ref{shc}) yields a spin Hall conductivity value of $0.067\frac{q}{4\pi}$ and the correction arising from the nonequilibrium distribution function  is estimated from (\ref {shc1}) as $0.030\frac{q}{4\pi}$.  Observe that they are of the same order, furthermore their combined effect yields a spin Hall conductivity value of $0.097\frac{q}{4\pi}$. Thus, we conclude that in principle the spin Hall conductivity in rotating coordinates should be observable. We can also make an estimate of the $\Omega$ and $B$ dependent terms in (\ref{jf02d}), the spin current density due to  equilibrium distribution function. We take $B=1 {\rm T}$, $\Omega=1 {\rm kHz} $ and $R=10 {\rm mm}$. The second term in (\ref{jf02d}) proportional to $\Omega B R$ yields a spin current approximately  $10^{-6} {\rm A/m}.$ For the same values, the estimated spin current in Ref.\cite{mism2} is roughly $10^{-8} {\rm A/m}.$ The difference arises from the  $\frac{\mu-m}{m}$ factor. Third term in    (\ref{jf02d}) proportional to $\Omega^2Rm$ yields a spin current around $10^{-13} {\rm A/m},$ which is negligible in comparison with the $\Omega BR$ term. 
  The structure of the three dimensional spin current densities are similar to the two dimensional ones.   However, our results are valid only at $T=0,$ because analytic calculations for $T\neq 0$   on general grounds are not available.   In the latter case one can try to compute spin current densities by treating the magnetic field and angular velocity as perturbations like the electric field. Obviously, one can  handle them   by  numerical analysis.   Nevertheless, in this work the necessary ingredients to construct the spin currents within the semiclassical approach are established.

\begin{center}
	\vspace{1cm}
	\noindent
	{\bf\Large{Acknowledgment}}
\end{center}

This work is supported by the Scientific and Technological Research Council of Turkey (T\"{U}B\.{I}TAK) Grant No. 115F108.

\vspace{1cm}
\noindent

\newcommand{\PRL}{Phys. Rev. Lett. }
\newcommand{\PRB}{Phys. Rev. B }
\newcommand{\PRD}{Phys. Rev. D }

\end{document}